\documentstyle[psfig,preprint,aps]{revtex}
\def\lsim{\:\raisebox{-0.5ex}{$\stackrel{\textstyle<}{\sim}$}\:}
\def\gsim{\:\raisebox{-0.5ex}{$\stackrel{\textstyle>}{\sim}$}\:}
\newcommand {\ignore}[1]{}

\renewcommand{\arraystretch}{1.5}
\newcommand{\noi}{\noindent}
\newcommand{\bc}{\begin{center}}
\newcommand{\ec}{\end{center}}

\def\ifmath#1{\relax\ifmmode #1\else $#1$\fi}

\def\3quarter{{\textstyle{3 \over 4}}}

\def\vs{\vskip}

\def\lf{\leaders\hbox to 1em{\hss.\hss}\hfill}
\def\21{$SU(2) \ot U(1)$}
\def\321{$SU(3) \ot SU(2) \ot U(1)$}
\def\ne{\hbox{$\nu_e$ }}
\def\nm{\hbox{$\nu_\mu$ }}

\def\eq#1{{eq. (\ref{#1})}}

\def\fig#1{{Fig. \ref{#1}}}

\def\lsim{\raise0.3ex\hbox{$\;<$\kern-0.75em\raise-1.1ex\hbox{$\sim\;$}}}
\def\gsim{\raise0.3ex\hbox{$\;>$\kern-0.75em\raise-1.1ex\hbox{$\sim\;$}}}

\def\beq{\begin{equation}}
\def\eeq{\end{equation}}
\def\bef{\begin{figure}}
\def\eef{\end{figure}}
\def\bet{\begin{table}}
\def\eet{\end{table}}
\def\bea{\begin{eqnarray}}
\def\ba{\begin{array}}
\def\ea{\end{array}}
\def\bi{\begin{itemize}}
\def\ei{\end{itemize}}
\def\ben{\begin{enumerate}}
\def\een{\end{enumerate}}

\def\ot{\otimes}
\def\eea{\end{eqnarray}}
\def\nps#1#2#3{        {\it Nucl. Phys. B (Proc. Suppl.) }{\bf #1} (19#2) #3}

\def\prl#1#2#3{          {\it Phys. Rev. Lett. }{\bf #1} (19#2) #3}

\def\n.c.#1#2#3{         {\it Nuovo Cim. }{\bf #1} (19#2) #3}
\def\r.n.c.#1#2#3{       {\it Riv. del Nuovo Cim. }{\bf #1} (19#2) #3}

\begin{document}

\draft

\preprint{\vbox{
\hbox{hep-ph/9801368}
\hbox{FTUV/97-72, IFIC/97-103}
\hbox{IFT - P.006/98} 
\hbox{January 1998}}
}
\title{Update on Atmospheric Neutrinos}
\author{M.\ C.\ Gonzalez-Garcia$^{1,2}$
\footnote{E-mail concha@evalvx.ific.uv.es}, 
H.\ Nunokawa$^1$
\footnote{E-mail nunokawa@flamenco.ific.uv.es}, 
O.\ L.\ G.\ Peres$^1$
\footnote{E-mail operes@flamenco.ific.uv.es},
T.\ Stanev$^3$ 
\footnote{E-mail stanev@bartol.udel.edu } and \\
J.\ W.\ F.\ Valle$^1$ 
\footnote{E-mail valle@flamenco.ific.uv.es}}
\address{$^1$ \it Instituto de F\'{\i}sica Corpuscular - C.S.I.C.\\
Departament de F\'{\i}sica Te\`orica, Universitat de Val\`encia\\
46100 Burjassot, Val\`encia, SPAIN \\
http://neutrinos.uv.es
}
\address{$^2$\it Instituto de F\'{\i}sica Te\'orica, 
             Universidade Estadual Paulista \\   
             Rua Pamplona 145,
             01405--900 S\~ao Paulo, BRAZIL} 
\address{$^3$\it Bartol Research Institute, University of Delaware, \\
Newark, Delaware 19716, USA.}
\maketitle
\begin{abstract}
\noi We discuss the impact of recent experimental
results on the determination of atmospheric neutrino oscillation
parameters.  We use all published results on atmospheric neutrinos,
including the preliminary large statistics data of Super-Kamiokande.
We re--analyze the data in terms of both $\nu_\mu \to \nu_\tau$ and
$\nu_\mu \to \nu_e$  channels using new improved calculations of the
atmospheric neutrino flux. We compare the sensitivity attained in
atmospheric neutrino experiments with those of accelerator and reactor
neutrino oscillation searches, including the recent Chooz
experiment. We briefly comment on the implications of atmospheric
neutrino data in relation to future searches for neutrino oscillations
with long baselines, such as the K2K, MINOS, ICARUS and NOE
experiments. 
\end{abstract}

\vskip 0.5cm



\section{Introduction}

Atmospheric neutrinos are produced in cascades initiated by collisions
of cosmic rays with the Earth's atmosphere.~\cite{review}.  Some of
the mesons produced in these cascades, mostly pions and kaons, decay
into electron and muon neutrinos and anti-neutrinos.
The predicted absolute fluxes of neutrinos produced by cosmic-ray
interactions in the atmosphere are uncertain at the 20\% level.  The
ratios of neutrinos of different flavour are however expected to be
accurate to better than 5\%.  Since $\nu_e$ is produced mainly from
the decay chain $\pi \to \mu \nu_\mu$ followed by $\mu \to e \nu_\mu
\nu_e$, one naively expects a $2:1$ ratio of $\nu_\mu$ to $\nu_e$. In
practice, however, the expected ratio of muon-like interactions to
electron-like interactions in each experiment is more uncertain
\cite{stanev,Barish}.

Several experiments have observed atmospheric neutrino interactions.
Two underground experiments, Kamiokande \cite{kamisub,kamimul} and
IMB \cite{IMB}, use water-Cerenkov detectors. These experiments have
detected a ratio of $\nu_\mu$-induced events to $\nu_e$-induced
events smaller than the expected one \cite{Barish}.  In particular
Kamiokande has performed separate analyses for both sub-GeV neutrinos
\cite{kamisub} and multi-GeV neutrinos \cite{kamimul}, which show the
same deficit. Although some of the experiments, such as Fr\'ejus
\cite{frejus} and NUSEX \cite{nusex}, have not found evidence for
this anomaly, and others, e.g. Soudan2, are not yet conclusive, the
recent Super-Kamiokande data provides a strong support for an
atmospheric muon neutrino deficit. This encourages us to reconsider
the analysis of atmospheric neutrino data from the point of view of
neutrino oscillation interpretation. The recent improved data sample
of Super-Kamiokande has now better statistics than achieved in the
whole Kamiokande phase.

We also include the new data of Soudan2 \cite{Soudan2} in our analysis.
These new data as well as the previous experimental
results are summarized in table \ref{tab:data}.  Here
$R_{\mu/e}/R^{MC}_{\mu/e}$ denotes the double ratio of
experimental-to-expected ratio of muon-like to electron-like
events. The expected ratio $R^{MC}_{\mu/e}$ is obtained by folding a
prediction for the atmospheric neutrino flux with the properties of
every individual detector through a Monte Carlo procedure.

Apart from studying the impact of the new data, our motivation for the
present reanalysis of atmospheric neutrino data is theoretical.  In
this regard, we first of all include the results of a recent
calculation of the atmospheric neutrino fluxes as a function of zenith
angle~\cite{agrawal}, including the muon polarization
effect~\cite{volkova}. Moreover, we develop an independent procedure
for the comparison of results from different experiments. We
demonstrate that our theoretical calculation of the energy
distribution of the event rates is in good agreement with the MC
expectations. The comparison of the experimental results presented
below thus reflects the significance of the atmospheric neutrino
anomaly and provides evidence for neutrino oscillations.

In this paper we analyze the impact of recent experimental results on
atmospheric neutrinos from Super-Kamiokande and Soudan2 on the
determinations of atmospheric neutrino oscillation parameters, both
for the $\nu_\mu \to \nu_\tau$ and $\nu_\mu \to \nu_e$ channels.  In
so-doing we take into account recent theoretical improvements in flux
calculations and neutrino-nucleon cross sections. The new
Super-Kamiokande data produce a downwards shift in the allowed
($\sin^2 2\theta, \Delta m^2)$ region, when compared with
pre-Super-Kamiokande results.  Nevertheless we show that the $\nu_\mu
\to \nu_e$ oscillation hypothesis is barely consistent with the recent
negative result of the Chooz reactor \cite{chooz}. The sensitivity
attained in atmospheric neutrino observations in the $\nu_\mu \to
\nu_\tau$ channel is also compared with those of accelerator neutrino
oscillation searches, for example at E776 and E531, as well as the
present CHORUS \cite{chorus} and NOMAD \cite{nomad} as well as the
future experiments being discussed at present.

\section{Atmospheric Neutrino Fluxes}

A contemporary calculation of the atmospheric neutrino fluxes consists
of a Monte Carlo procedure that folds the measured energy spectra and
chemical composition of the cosmic ray flux at the top of the
atmosphere with the properties of hadronic interaction with the light
atmospheric nuclei. Since the properties of the secondary mesons are
extremely well known, the accuracy of the calculation is determined by
the uncertainty of the two sets of assumptions -- about the primary
cosmic ray flux and about the hadronic interactions on light nuclei.

In order to avoid the uncertainty in the absolute magnitude of the
cosmic ray flux experiments measure the ratio of electron to muon
neutrinos, which is very stable in different calculations. The
absolute normalization of the atmospheric neutrino flux is still very
important for the interpretation of the observed muon neutrino
deficit. If it turns out that the measured numbers of electron--like
interactions agrees with the predictions and there is an absolute
deficit of muon--like interactions, the causes must be $\nu_\mu$
disappearance. If experiments measure the right amount of $\nu_\mu$
and an excessively large number of $\nu_e$ there must be a reason for
$\nu_e$ appearance, such as $\nu_\mu \to \nu_e$ oscillations or a
background process that generates $\nu_e$ or $e^\pm$ events in the
detectors.
 
We use the new neutrino flux calculations of the Bartol
group~\cite{agrawal} which are performed with an updated version of
the TARGET interaction event generator. The cosmic ray flux model is
discussed in detail in Ref.~\cite{agrawal}. The treatment of the
hadronic collisions is very similar to that in earlier
calculations~\cite{BGS}. There are only minor improvements in the
treatment of the resonant region for low energy collisions and in the
cross section for production of $\Lambda K$ pairs above 1000 GeV. The
improvements in the low energy (2 -- 3 GeV) range affects slightly the
fluxes of 100 -- 300 MeV neutrinos, while the kaon spectra at high
energy change the neutrino to anti-neutrino ratios above 100 GeV and
thus mostly affect the predictions for the flux of upward going
neutrino induced muons.

In the absence of geomagnetic effects the fluxes of GeV neutrinos are
practically the same as those of Ref.~\cite{BGS}. Much more
significant difference is introduced by the improved treatment of the
geomagnetic effects~\cite{LSgeomag}. The probability for low rigidity
cosmic rays to penetrate to the atmosphere and to produce neutrinos is
calculated using a realistic model of the geomagnetic field and
accounting for the shadow of the Earth. As a result the neutrino
fluxes at experimental locations with high geomagnetic cutoff, such as
Kamioka, are significantly lower than in Ref.~\cite{BGS}. At high
geomagnetic latitude the new fluxes are comparable to the original
ones.

This new set of fluxes, as well as the original one, belongs to a
group (together with the calculation of Ref.~\cite{HKHM}) of
atmospheric neutrino flux predictions of relatively high
magnitude. The expected magnitude of the atmospheric neutrinos was
discussed by the authors of different predictions~\cite{Getall} who
identified the reason for the differences in the treatment of the
nuclear target effect in the hadronic collisions in the
atmosphere. Calculations that assume that pion multiplicities in {\it
pp} and {\it pAir} collisions are similar~\cite{BN} predict low
neutrino flux magnitude. The event generator TARGET produces pion
multiplicity that is higher by a factor of $\sim$1.6 in {\it pAir}
interactions above the resonant region.

The muon fluxes at different atmospheric depths generated with the
same code as the new neutrino fluxes were compared to the measurements
of the MASS experiment~\cite{Circella}. The predicted altitude profile
of muons of energy above 1 GeV agrees with the measured one extremely
well.

\section{Atmospheric Neutrino Cross Sections and Event Distributions}

For each experiment the expected number of $\mu$-like and $e$-like
events, $N^0_\alpha$, $\alpha = \mu, e$, in the absence of
oscillations can be computed as
\begin{equation}
N^0_{\alpha}= \int \frac{d^2\Phi_\alpha}{dE_\nu d(\cos\theta_\nu)}
\kappa_\alpha(h,\cos\theta_\nu,E_\nu)
\frac{d\sigma}{dE_\alpha}\varepsilon(E_\alpha)
dE_\nu dE_\alpha d(\cos\theta_\nu) dh 
\; .
\label{event0}
\end{equation}
Here $E_\nu$ is the neutrino energy and $\Phi_\alpha$ is the flux of
atmospheric neutrinos of type $\alpha=\mu ,e$; $E_\alpha$ is the final
charged lepton energy and $\varepsilon(E_\alpha)$ is the detection
efficiency for such charged lepton; $\sigma$ is the neutrino-nucleon
interaction cross section, $\nu \; N \to N'\; \ell$; $\theta_\nu$ is
the angle between the vertical direction and neutrinos
($\cos\theta_\nu$=1 corresponds to the downward direction).  For some
experiments, such as Fr\'ejus, we also include neutral current events
which are misidentified as charged current ones. In \eq{event0}
$\kappa_\alpha$ is the distribution of $h$ which is the slant distance
from the production point to the sea level for $\alpha$ type neutrinos
with energy $E_\nu$ and zenith angle $\theta_\nu$.  We took the
distribution from ref. \cite{pathlength} which is normalized as
\begin{equation}
\int \kappa_\alpha(h,\cos\theta_\nu,E_\nu) dh = 1.
\end{equation}

As discussed in Sec. II, the neutrino fluxes, in particular in the
sub-GeV range, depend on the solar activity.  In order to take this
fact into account we use in \eq{event0} and also in sec. IV the
averaged neutrino flux defined as follows,
\begin{equation}
{\Phi}_\alpha \equiv 
c_{max} \Phi_\alpha^{max} + c_{min} \Phi_\alpha^{min}, 
\end{equation}
where $\Phi_\alpha^{max}$ and $\Phi_\alpha^{min}$ are the atmospheric
neutrino fluxes when the sun is most active (solar maximum) and quiet
(solar minimum), respectively.  The coefficients $c_{max}$ and
$c_{min}$ ($=1-c_{max}$) are determined according to the running
period of each experiment assuming that the flux changes linearly with
time between solar maximum and minimum.  This is a first order
correction for the solar modulation of the primary cosmic ray flux
which has not been included in previous analyses.

\subsection{Cross Sections}

\begin{figure}
\centerline{\protect\hbox{
\psfig{file=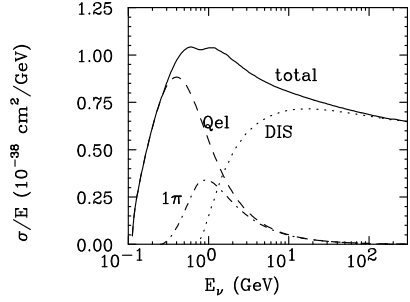,height=7cm,width=9cm,angle=90}
}}
\vglue -1.5cm
\caption{Neutrino-nucleon cross section used in this paper. }
\label{sigma} 
\end{figure}

In order to determine the expected event rates for the various
experiments we use the neutrino-nucleon cross sections presented in
\fig{sigma}.  We consider separately the contributions to the cross
section from the exclusive channels of lower multiplicity,
quasi-elastic scattering and single pion production, and include all
additional channels as part of the deep inelastic (DIS) cross section
\cite{paolo}:
\begin{equation} 
\sigma^{CC}=\sigma_{QE}+\sigma_{1\pi}+\sigma_{DIS}\; .
\end{equation}
The quasi-elastic cross section for a neutrino with energy $E_\nu$ 
is given by \cite{smith}:
\begin{equation}
\frac{d\sigma_{QE}}{d|q^2|}(\nu n \to  \ell^- p)= 
\frac{M^2 G_F^2 cos^2\theta_c}
{8 \pi E_\nu^2} \left[A_1(q^2)-A_2(q^2) \frac{s-u}{M^2} +A_3(q^2) 
\frac{(s-u)^2}{M^4}
\right]
\end{equation}
where $s-u=4 M E_\nu +q^2 -m_\ell^2$, $M$ is the proton mass, $m_\ell$ is the 
charged lepton mass and $q^2$ is the momentum transfer.  
For $\nu p \to  \ell^+ n$, the same formula applies with the change
$A_2 \to  -A_2$.  The functions $A_1$, $A_2$, and $A_3$ can be written in
terms of axial and vector form factors:
\begin{equation}
\begin{array}{ll}
A_1=&\frac{\displaystyle m_\ell^2-q^2}{\displaystyle 4M^2} 
\Bigl[\bigl(4-\frac{\displaystyle q^2}{\displaystyle M^2}\bigr)|F_A|^2
-\bigl(4 +\frac{\displaystyle q^2}{\displaystyle M^2}\bigr)|F_V^1|^2
-\frac{\displaystyle q^2}{\displaystyle M^2}|\xi F_V^2|^2
-\frac{\displaystyle 4 q^2}{\displaystyle M^2}
\mbox{Re}(F_V^{1\star}\xi F^2_V) \\
& -\frac{\displaystyle m_\ell^2}{\displaystyle M^2}\left(|F_V^1 + \xi F_V^2|^2+
|F_A|^2\right) \\
A_2= & -\frac{\displaystyle q^2}{\displaystyle M^2}
\mbox{Re} [F_A^\star (F_V^1 + \xi F_V^2)] \\
A_3=&  -\frac{\displaystyle 1}{\displaystyle 4}\left(|F_A|^2+|F_V^1|^2 -
\frac{\displaystyle q^2}{\displaystyle 4M^2}|\xi F_V^2|^2 \right)
\end{array}
\end{equation} 
where we have neglected second order currents and we have assumed CVC.
With this assumption all form factors are real and can be written as
\begin{equation}
\begin{array}{l}
F_V^1(q^2)= \left(1-\frac{\displaystyle q^2}{\displaystyle 4 M^2}\right)^{-1}
\left(1-\frac{\displaystyle q^2}{\displaystyle M_V^2}\right)^{-2}
\left[1-\frac{\displaystyle q^2}{\displaystyle 4 M^2}(1+\mu_p-\mu_n)\right]\\
\xi F_V^2(q^2)= \left(1-\frac{\displaystyle q^2}{\displaystyle 4 M^2}\right)^{-1}
\left(1-\frac{\displaystyle q^2}{\displaystyle M_V^2}\right)^{-2}
(\mu_n-\mu_p)\\
F_A=F_A(0)\left(1-\frac{\displaystyle q^2}{\displaystyle 4 M_A^2}\right)^{-2} \; . 
\end{array}
\end{equation}
$\mu_p$ and $\mu_n$ are the proton and neutron anomalous magnetic
moments and the vector mass, $M_V^2=0.71$ GeV$^2$, is measured with
high precision in electron scattering experiments. The largest
uncertainties in this calculation are associated with the axial form
factor. In our simulation we use $F_A(0)=-1.23$ which is known from
neutron beta decay. The axial mass used by the different
collaborations varies in the range $M_A^2=0.71-1.06$ GeV$^2$.

So far we have neglected nuclear effects. The most important of such
effects is due to the Pauli principle. Following Ref. \cite{smith} we
include it by using a simple Fermi gas model. In this approximation
the cross section of a bound nucleon is equal to the cross section of
a free nucleon multiplied by a factor $(1-N^{-1}D)$.  For neutrons
\begin{equation}
\begin{array}{lr}
D=Z &\  \mbox{for   } 2z \leq u-v \\
D=\frac{1}{2} A\left[1-\frac{3z}{4} (u^2+v^2)+\frac{z^3}{3}+\frac{3}{32 z}
(u^2-v^2)^2\right] &\ \mbox{for } u-v \leq 2z \leq u+v \\
D=0 &\  \mbox{for   } 2z \geq u+v 
\end{array}
\end{equation}
with $z= \Big[\sqrt{(q^2+m_\ell^2)^2/(4M^2)-q^2}\:\:\: \Big] /(2 k_f^2)$, $u=(2
N/A)^{1/3}$, and $v=(2 Z/A)^{1/3}$. 
Here $A,Z,N$ are the nucleon, proton and neutron numbers 
and $k_f$ is the Fermi momentum,
$k_f=0.225$($0.26$), for oxygen (iron).  For protons, the same formula
applies with the exchange $N\leftrightarrow Z$. The effect of this
factor is to decrease the cross section. The decrease is larger for
smaller neutrino energy. For energies above 1 GeV the nuclear effects
lead to an 8 \% decrease on the quasi-elastic cross section.

For single pion production we use the model of Fogli and Nardulli 
\cite{cross} which includes hadronic masses below $W=1.4$ GeV. 
Deep inelastic cross sections are usually described in terms of
the variables $y=1-E_{\ell}/E_\nu$ and $x=-q^2/(2 M E_\nu y)$. In the
parton model
\begin{equation}
\frac{d\sigma_{DIS}}{dx dy}
\left(\begin{array}{l}\nu \\[-0.2cm]
                      \bar\nu 
      \end{array}\right)
=\frac{G_F^2 s x}{4 \pi}
\left[F_1\mp F_3 +(F_1\pm F_3)(1-y)^2\right] 
\end{equation}
where $F_1$ and $F_3$ are given in terms of the parton distributions.
For isoscalar targets $F_1=2{\displaystyle \sum_i} (q_i+\bar q_i)$ 
and $F_3={\displaystyle \sum_i} (\bar q_i -q_i)$.  
In order to avoid double counting we follow the approach of Ref. \cite{paolo}
and we integrate the deep inelastic contribution in the region 
$W>W_c$ which implies $2 M E_\nu y (1-x) \geq W_c^2 -M^2$
where $W_c = 1.4$ GeV.

The final necessary ingredient are the detector efficiencies given by
the experiments.  These are, in general, functions of the incident
neutrino energy and the detected lepton energy and flavour.  We took
these efficiencies from refs. \cite{kajita} for Kamiokande sub-GeV and
IMB, \cite{frejus} for Fr\'ejus, \cite{nusex} for Nusex. The
efficiencies for the Kamiokande multi-GeV and $e$-like events for
Soudan2 are provided by the experimentalists and for Soudan2
$\mu$-like events the efficiencies are determined in such a way that
the energy distributions are well reproduced.  For the
Super-Kamiokande we are making some approximation based on the
information provided also by the experimentalists, as discussed
below.

\subsection{Event Distributions}

\begin{figure}
\centerline{\protect\hbox{\psfig{file=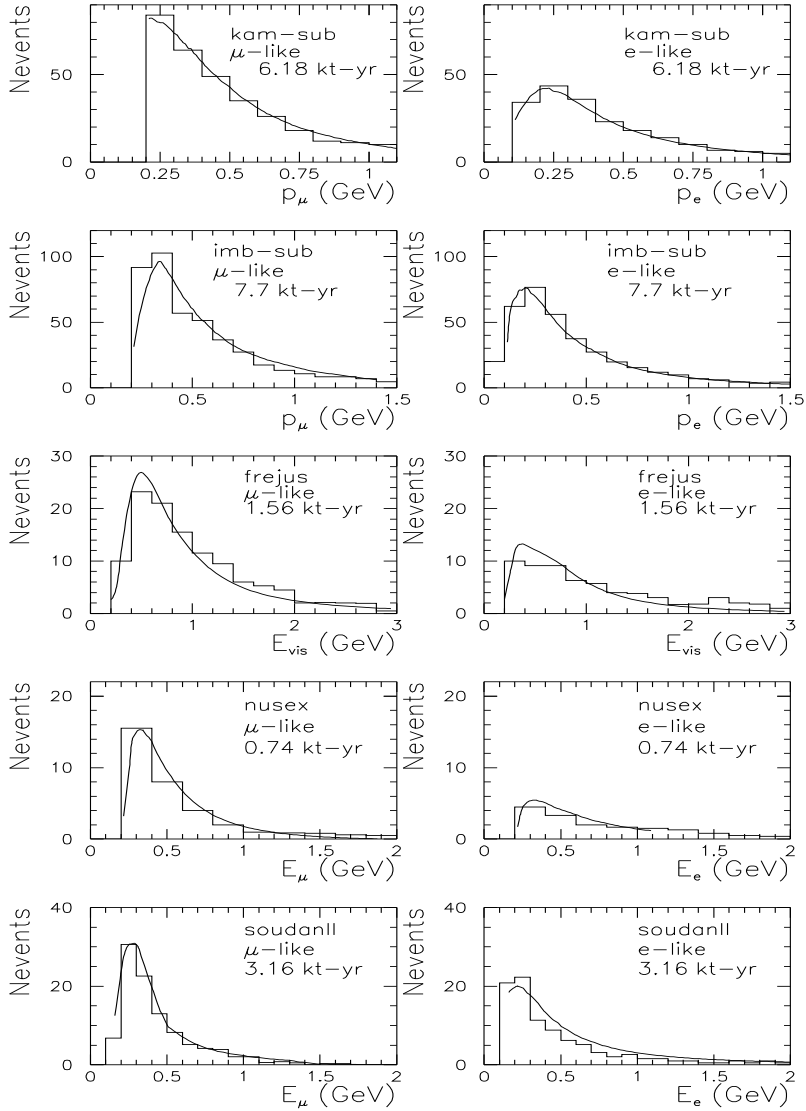,height=15.5cm,width=.9\textwidth,angle=0}}}
\caption{Expected energy distribution of Sub-GeV events (histogram) compared 
with our prediction (full line).}
\label{Nevsub} 
\end{figure}

In order to verify the quality of our simulation we compare our
predictions for the energy distribution of the events with the Monte
Carlo simulations of the different experiments in absence of
oscillation. In \fig{Nevsub} we show our predictions superimposed with
those from the experimental Monte Carlo for the sub-GeV
experiments, Kamiokande sub-GeV \cite{kamisub}, IMB \cite{IMB},  
Frejus \cite{frejus}, Nusex \cite{nusex} and Soudan II \cite{Soudan2}.
We can see that the agreement is very good.  
No additional normalization of the event rates has been
performed.

\begin{figure}
\centerline{\protect\hbox{\psfig{file=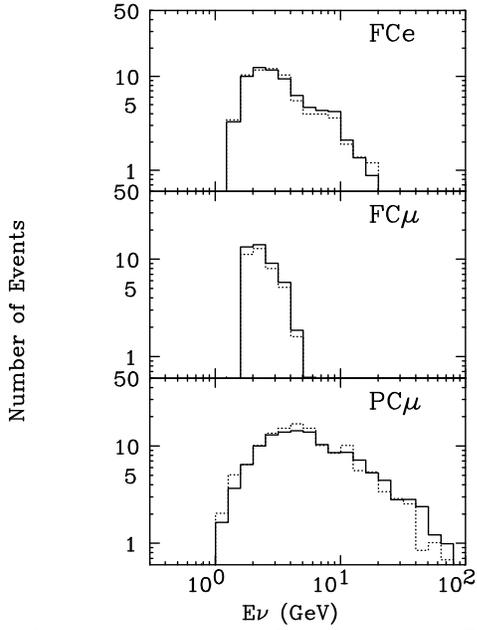,width=.7\textwidth,angle=90}}}
\caption{Expected neutrino energy distribution of Kamiokande Multi-GeV events 
(dashed histogram) compared with our prediction (full histogram) . }
\label{Nevmulti} 
\end{figure}

Similarly, in \fig{Nevmulti} we show the distribution of the fully
contained electron-like events and fully and partially contained muon
events for the Kamiokande multi-GeV sample compared with the
experimental Monte Carlo prediction given in Fig.2 of Ref
\cite{kamimul}.  Some comments are due.  In order to obtain these
distributions we have used detailed experimental efficiencies of
Kamiokande for detecting fully contained and partially contained
electron and muon events \cite{kajita}.  One must take into account
that the Monte Carlo distributions given in Fig.2 of Ref
\cite{kamimul} were generated using the fluxes of Honda {\sl et
al.}\cite{HKHM} while we are using the fluxes of Gaisser {\sl et al.}
\cite{BGS}.  Thus we have an absolute prediction for the number of
events for Kamiokande multi-GeV data and for their energy distribution
which is obtained under the same assumptions for the cross sections
and neutrino fluxes than the other sub-GeV experiments. For the sake
of comparison we also show in \fig{ang} (upper two panels) the angular
distribution of the events for Kamiokande multi-GeV in the absence of
oscillations as obtained from our calculation.

\begin{figure}
\centerline{\protect\hbox{\psfig{file=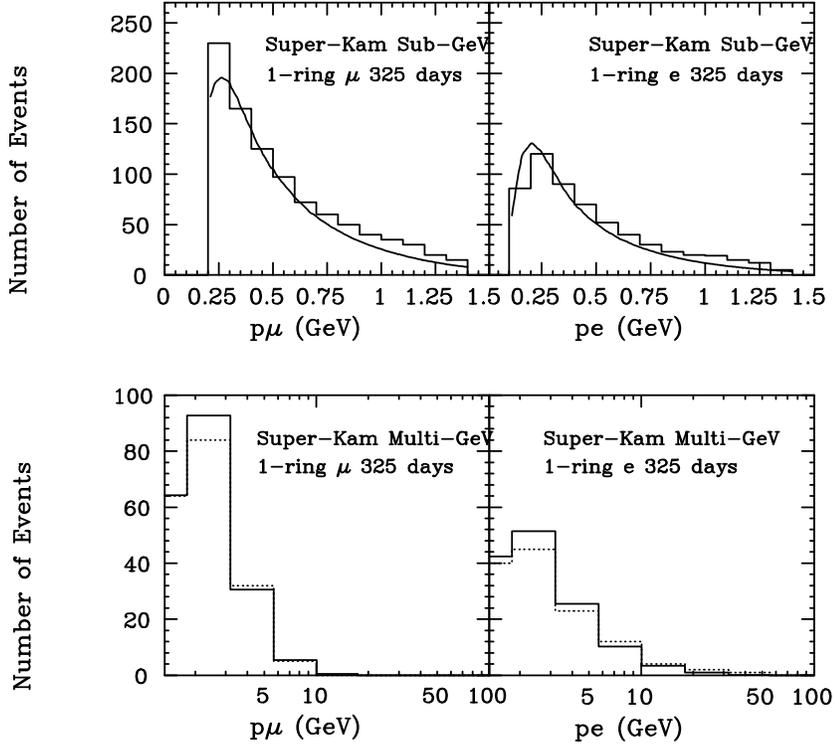,width=.9\textwidth,angle=90}}}
\caption{Expected energy distribution of Super-Kamiokande events. 
For the sub-GeV events the histogram represents the MC expectation 
while the full line is our prediction. For the Multi-GeV events 
the full histogram is our result while the dashed histogram gives the 
MC prediction. Both  our prediction and the MC prediction are 
based on the same flux calculations \protect\cite{agrawal}.}
\label{Nevsk} 
\end{figure}

\begin{figure}
\centerline{\hskip -1cm
\psfig{file=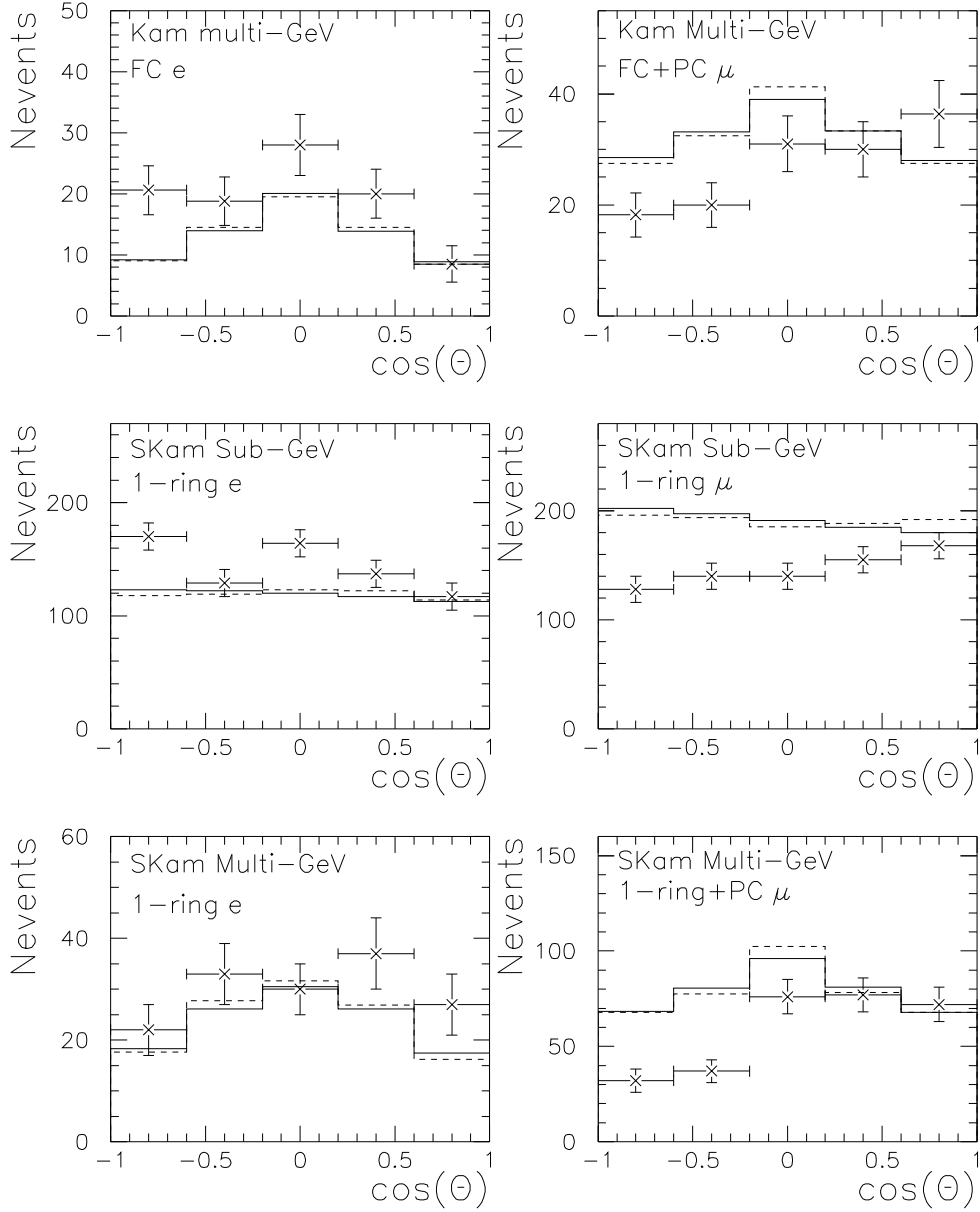,width=.9\textwidth}}
\caption{Expected angular distribution of Kamiokande multi-GeV events
and Super-Kamiokande events (dashed histogram) obtained by Monte Carlo
simulation by the experimental group compared with our predictions
(full histogram) and the experimental data. We note that in these
figures the MC prediction is based on Honda {\it et al.} fluxes
\protect\cite{HKHM} whereas ours is based on Bartol fluxes
\protect\cite{agrawal} normalized to the total number of expected
events with the Honda MC.}
\label{ang}  
\end{figure}

In \fig{Nevsk} we plot as in \fig{Nevsub} the expected energy
distribution also for Super-Kamiokande sub-GeV and multi-GeV data.
We also plot in \fig{ang} the angular distribution for
Super-Kamiokande sub-GeV (middle two panels) and multi-GeV data (lower
two panels).  We have used, as an approximation, the preliminary
acceptances \cite{inoue} of Super-Kamiokande for 325.8 days for fully
contained, 293 days for partially contained events as detection
efficiencies for final leptons, for sub-GeV as well as multi-GeV
data. In order to obtain the angular distribution of expected events
for in the multi-GeV range we have assumed that the lepton direction
is the same as the incident neutrino direction. 
%
Actually for the Kamiokande multi-GeV data, the average angle between
the incident neutrino and the lepton direction is about $15^\circ$. In
our calculation we have simulated this difference by smearing the
angular distribution with a Gaussian distribution with one-sigma width
of $15^\circ$.  As seen in \fig{ang} the effect of this approximation
is small.  At this point it is worth noting that the angular
distribution for multi-GeV electrons in the Super-Kamiokande sample is
flatter than in the Kamiokande data. The main reason for this
zenith-angle shape difference is due to the smaller selection
efficiency for the 1-ring e-like events at high energy in the
super-Kamiokande analysis \cite{kajita}. As a result the mean neutrino
energy was shifted to lower value and the mean angle between the
incident neutrino and the lepton direction becomes larger.  We have
simulated this effect by increasing the one-sigma width of the
smearing Gaussian to $25^\circ$ for the super-Kamiokande multi-GeV
electrons which effectively flattens the angular distribution as seen
in \fig{ang}.

On the other hand, for events in the sub-GeV range we have carefully taken 
into account the difference between the incoming neutrino angle and the
detected charged lepton scattering angle which is a function of the
incoming neutrino energy. As can be seen in \fig{ang} this leads to a
much flatter expected angular distribution for the sub-GeV neutrinos,
in agreement with the prediction from the experimental MC.

We also estimate the expected the ratio in the absence of oscillation
as
\begin{equation}
R^0_{\mu/e}= \frac{N^0_{\mu}}{N^0_{e}}\; ,
\end{equation}
where $N^0_{\mu}$ and $N^0_{e}$ are computed by \eq{event0}.  In Table
\ref{tab:our} we present our prediction for the expected ratio in the
absence of oscillations for the various experiments and compare it
with the expected MC results
\cite{kamisub,kamimul,IMB,frejus,nusex,Soudan2}.  Table \ref{tab:our}
also displays our prediction for the expected ratio for the Kamiokande
multi-GeV and Super-Kamiokande zenith angle distribution. We see that
the agreement between $R^{MC}_{\mu/e}$ and our prediction
$R^0_{\mu/e}$ are very good for most of the experiments.

\section{Atmospheric Neutrino Data Fits} 

We now consider the simplest interpretation of the atmospheric
neutrino anomaly in terms of the neutrino oscillation hypothesis.  For
definiteness we assume a two-flavor oscillation scenario, in which the
$\nu_\mu$ oscillates into another flavour either $\nu_\mu \to \nu_e$
or $\nu_\mu \to \nu_\tau$.

\subsection{Data Analysis Procedure}

In the presence of two-flavour neutrino oscillations, the expected
number of $\mu$ and $e$-like events, $N_\alpha$, $\alpha = \mu, e$ is
given by
\begin{equation}
N_\mu=N^0_{\mu\mu} \langle P_{\mu\mu}\rangle +N^0_{e\mu} \
\langle P_{e\mu}\rangle \; ,  \;\;\;\;\;\
N_e=N^0_{ee} \langle P_{ee}\rangle +N^0_{\mu e} \langle P_{\mu e}\rangle \; ,
\label{eventsnumber}
\end{equation}
where
\begin{equation}
N^0_{\alpha\beta}=\int \frac{d^2\Phi_\alpha}{dE_\nu d(\cos\theta_\nu)}
\kappa_\beta(h,\cos\theta_\nu,E_\nu)
\frac{d\sigma}{dE_\beta}\varepsilon(E_\beta)
dE_\nu dE_\beta d(\cos\theta_\nu)dh
\end{equation}
and
\begin{equation}
\langle P_{\alpha\beta}\rangle =\frac{1}{N^0_{\alpha\beta}}\int
\frac{d^2\Phi_\alpha}{dE_\nu d(\cos\theta_\nu)} 
\kappa_\beta(h,\cos\theta_\nu,E_\nu)
P_{\alpha\beta}
\frac{d\sigma}{dE_\beta}\varepsilon(E_\beta)
dE_\nu dE_\beta d(\cos\theta_\nu)dh\; .
\label{paverage}
\end{equation}
Here $P_{\alpha\beta}$ is the oscillation probability of 
$\nu_\beta \to \nu_\alpha$ for given values of 
$E_{\nu_\beta}, \cos\theta_\nu$ and $h$, i.e., 
$ P_{\alpha\beta} \equiv 
P(\nu_\beta \to \nu_\alpha; E_{\nu_\beta}, \cos\theta_\nu, h) $

We note that for the $\nu_\mu \to \nu_e$ channel, Earth matter effects
lead to oscillation probabilities which are different for neutrinos
and anti-neutrinos. Therefore, we separately compute $P_{\alpha\beta}$
in \eq{paverage} in order to correctly estimate the expected number of
events in each experiment (see sec. IV.B for more discussion).

When combining the results of the experiments we do not make use of
the double ratio, $R_{\mu/e}/R^{MC}_{\mu/e}$, 
but instead we treat the $e$ and $\mu$-like data separately, 
taking into account carefully the
correlation of errors. Following ref. \cite{fogli2} we define $\chi^2$
as
\begin{equation}
\chi^2 \equiv \sum_{I,J}
(N_I^{data}-N_I^{theory}) \cdot 
(\sigma_{data}^2 + \sigma_{theory}^2 )_{IJ}^{-1}\cdot 
(N_J^{data}-N_J^{theory}),
\label{chi2}
\end{equation}
where $I$ and $J$ stand for any combination of experimental data set
and type of events considered, i.e, $I = (A, \alpha)$ and $J = (B,
\beta)$ where, $A,B$ = Fr\'ejus, Kamiokande sub-GeV, IMB,... and
$\alpha, \beta = e,\mu$.  In \eq{chi2} $N_I^{theory}$ is the predicted
number of events calculated by \eq{eventsnumber} whereas $N_I^{data}$
is the number of observed events.  In \eq{chi2} $\sigma_{data}^2$ and
$\sigma_{theory}^2$ are the error matrices containing the experimental
errors and the MC errors respectively. They can be written as
\begin{equation}
\sigma_{IJ}^2 \equiv \sigma_\alpha(A)\, \rho_{\alpha \beta} (A,B)\,
\sigma_\beta(B),
\end{equation}
where $\rho_{\alpha \beta} (A,B)$ stands for the correlation between
the $\alpha$-like events in the $A$-type experiment and $\beta$-like
events in $B$-type experiment, whereas $\sigma_\alpha(A)$ and
$\sigma_\beta(B)$ are the errors for the number of $\alpha$ and
$\beta$-like events in $A$ and $B$ experiments, respectively. The
dimension of the error matrix varies depending on the combination of
experiments included in the analysis.  For each individual experiment,
the error matrix has dimension $2\times 2$ whereas for the full
experimental data set with binning (20 data for each flavor) its
dimension is $40\times 40$.

With this procedure of treating separately the $e$-like and $\mu$-like
data with correlation of errors, we avoid the non-Gaussian
nature of double ratio, as pointed out in ref. \cite{fogli2}.

We compute $\rho_{\alpha \beta} (A,B)$ as in ref. \cite{fogli2}.  A
detailed discussion of the errors and correlations used in our
analysis can be found in the Appendix. In Table \ref{tab:chi2} we show
the values of $\chi^2$ and the confidence level in the absence of
oscillation. In our analysis, we have conservatively assumed 30\%
uncertainty regarding to the absolute neutrino flux, in order to
generously account for the spread of neutrino flux predictions in
different calculations \footnote{For a brief discussion of the effect
of the assumed flux uncertainties, see section V}.

Next we minimize the $\chi^2$ function in \eq{chi2} and determine the
allowed region in the $\sin^22\theta-\Delta m^2$ plane, for a given
confidence level, defined as,
\begin{equation}
\chi^2 \equiv \chi_{min}^2  + 4.61\ (9.21)\   \  
\  \mbox{for}\  \ 90\  (99) \% \ \  \mbox{C.L.}
\end{equation}

\subsection{$\nu_\mu \to  \nu_e$ channel }

\begin{figure}
\centerline{\protect\hbox{\psfig{file=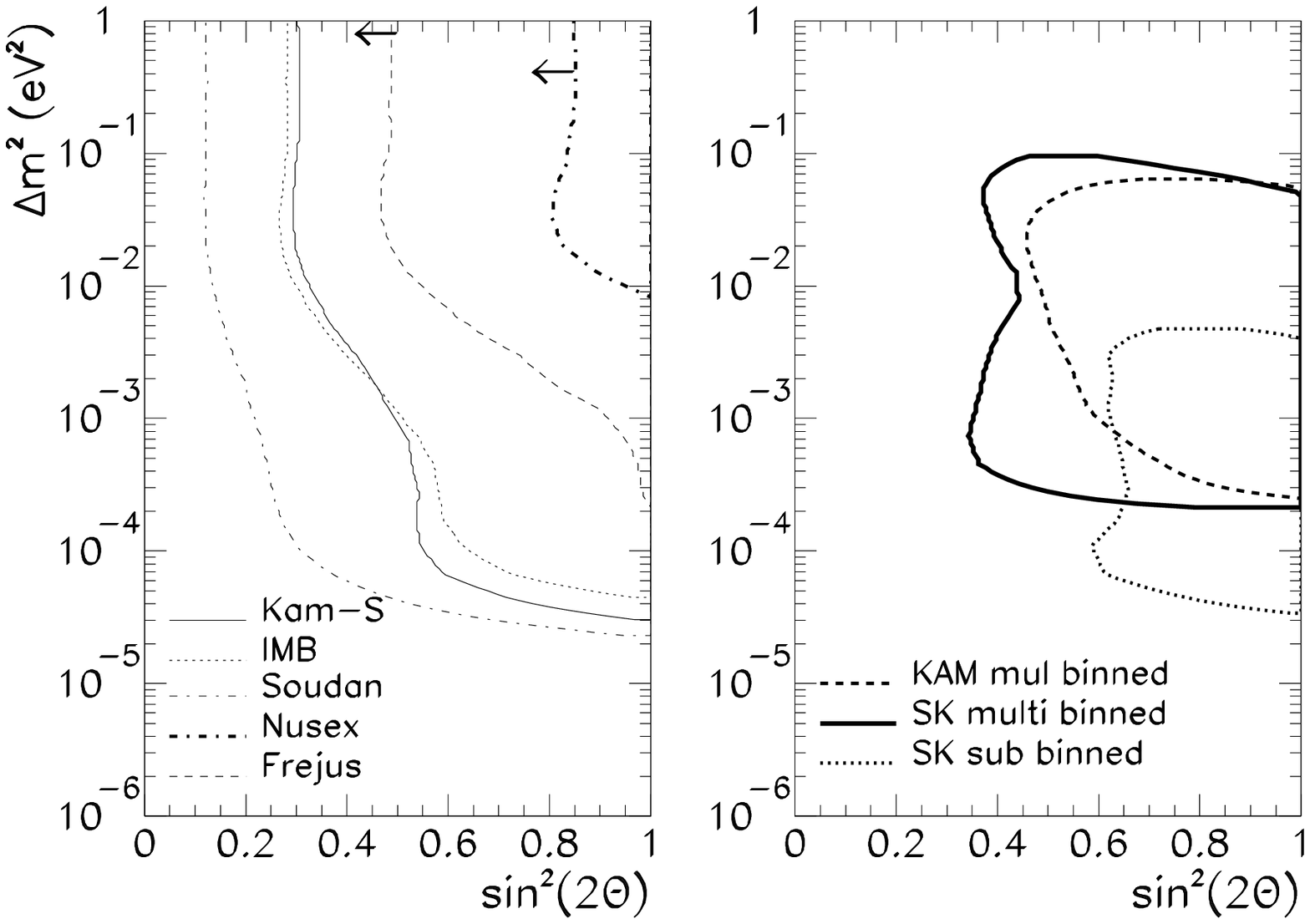,width=.75\textwidth}}}
\caption{Allowed $\nu_\mu \to  \nu_e$ oscillation parameters at 90 \% CL
for each individual experiment neglecting Earth matter effects.}
\label{muevacie} 
\end{figure}

The results of our $\chi^2$ fit of atmospheric neutrino data obtained
at the various individual water-Cerenkov and iron calorimeter
detectors for the $\nu_\mu \to \nu_e$ channel are shown in
\fig{muevacie}.  The left-pointing arrows in the left figure in
\fig{muevacie} correspond to the negative results of Frejus and
Nusex. So far we have not included in the above analysis the
constraints that arise from the inclusion of the angular dependence of
the data in the Kamiokande multi-GeV data as well as the
Super-Kamiokande data. In the right hand panel of \fig{muevacie} we
show how the binned results of Kamiokande and Super-Kamiokande give
rise to a region of oscillation parameters that cuts out the large
$\Delta m^2$ values.  Moreover one can see that the Super-Kamiokande
binned sub-GeV data yield a somewhat lower value of $\Delta m^2$ than
the multi-GeV data.

\begin{figure}
\centerline{\protect\hbox{\psfig{file=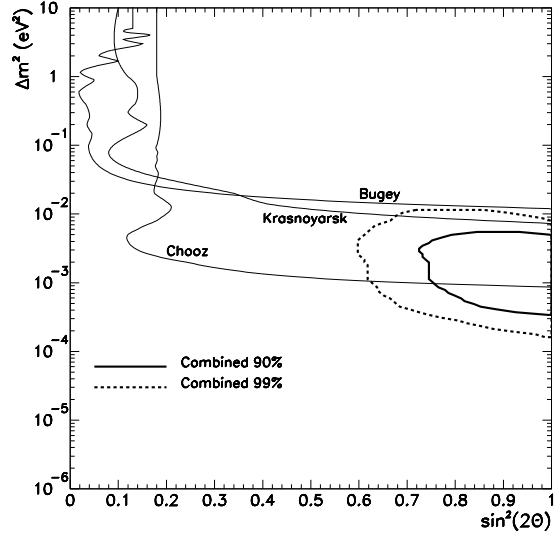,width=0.5\textwidth}}}
\caption{Allowed $\nu_\mu \to  \nu_e$ oscillation parameters
for all experiments combined at 90 and 99 \% CL neglecting 
Earth matter effects. For comparison we also plot the presently 
excluded region from reactor experiments. }
\label{muevacall} 
\end{figure}

The effects of combining all atmospheric neutrino data from the
various experiments for the $\nu_\mu \to \nu_e$ channel are shown in
\fig{muevacall}. This figure show the allowed $\nu_\mu \to \nu_e$
oscillation parameters for all experiments combined at 90 and 99 \%
CL. For comparison we have also plotted in \fig{muevacall} the
presently excluded region from reactor experiments, Krasnoyarsk
\cite{krasnoyarsk}, Bugey \cite{bugey} and the recent Chooz
long-baseline result \cite{chooz}.

We have so far neglected Earth matter effects \cite{msw}, both in
\fig{muevacie} and \fig{muevacall}.  In order to take into account the
earth matter effect in our analysis we have separately computed, by
numerical integration, the oscillation probabilities, $P(\nu_\mu \to
\nu_e) = P(\nu_e \to \nu_\mu)$ and $P(\bar{\nu}_\mu \to \bar{\nu}_e) =
P(\bar{\nu}_e\to \bar{\nu}_\mu)$.  This is necessary, since the matter
effect distinguishes neutrinos from anti-neutrinos. We have used the
approximate analytic expression for the electron density profile in
the Earth obtained in ref. \cite{lisi}.  In order to save computation
(CPU) time we have neglected the matter effect for neutrino
oscillation parameters in the range,
\begin{equation}
\frac{\Delta m^2}{E} >
10^{-11} \mbox{eV},
\label{neglect}
\end{equation}
since the maximum value of the matter potential (at the Earth center)
is at most,
\begin{equation}
V_{matter} \sim  10^{-12} \mbox{eV},
\end{equation}
and the matter effect on the probability is small if condition
\eq{neglect} is satisfied.

\begin{figure}
\centerline{\protect\hbox{\psfig{file=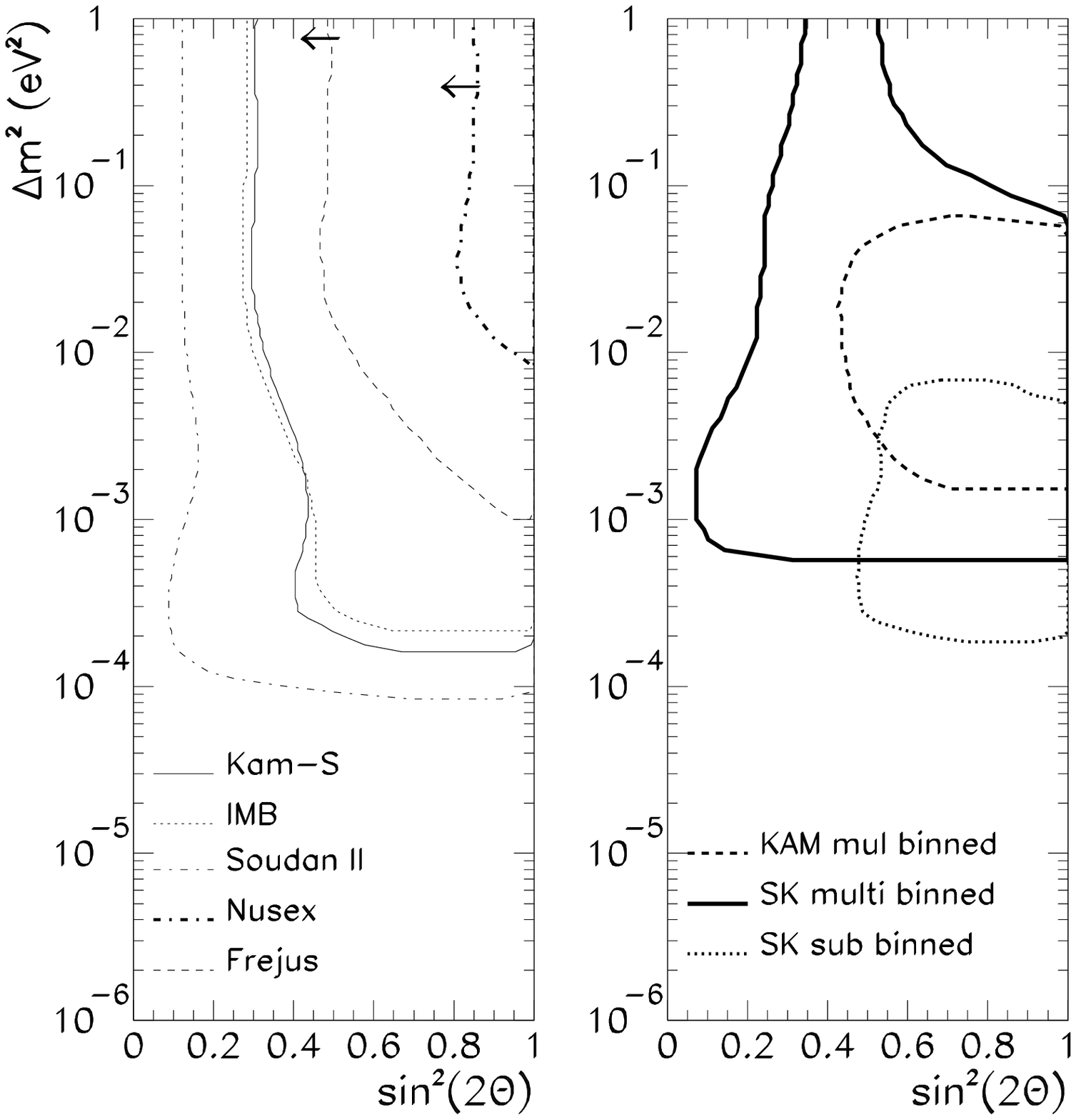,width=0.75\textwidth}}}
\caption{Allowed $\nu_\mu \to  \nu_e$ oscillation parameters at 90\% CL
 for each individual experiment including Earth matter effects.}
\label{muematie} 
\end{figure}

\begin{figure}
\centerline{\protect\hbox{\psfig{file=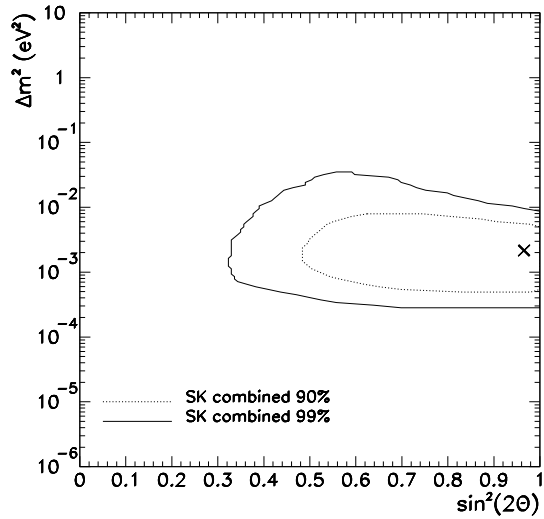,width=0.5\textwidth}}}
\caption{Allowed $\nu_\mu \to \nu_e$ oscillation parameters for
Superkamiokande experiment combined at 90 and 99 \% CL including Earth
matter effects.}
\label{muematsk} 
\end{figure}

In \fig{muematie} we show the allowed $\nu_\mu \to \nu_e$ oscillation
parameters for each individual experiment including Earth matter
effects. Unlike the previous case where matter effects were neglected,
a noticeable new feature in this case is that the Super-Kamiokande
multi-GeV data now allows large $\Delta m^2$ values, even if binning
is taken into account. The allowed $\nu_\mu \to \nu_e$ oscillation
parameters for the Super-Kamiokande binned data combined at 90 and 99
\% CL including Earth matter effects in shown in \fig{muematsk}. An
interesting feature to note here is that by adding the matter effects
the allowed regions lie higher in $\Delta m^2$ than when matter
effects are neglected.  This is because for smaller $\Delta m^2$,
i.e. when $\Delta m^2 \cos 2\theta /2E$ is much smaller than
$V_{matter}$ the effective conversion amplitude $\sin ^22 \theta_m$
where $\theta_m$ is the mixing angle in matter, is smaller than that
of the vacuum one, i.e. $\sin ^22 \theta$. In other words, in this
region matter suppresses the conversion and it becomes harder to fit
for the atmospheric neutrino anomaly.

\begin{figure}
\centerline{\protect\hbox{\psfig{file=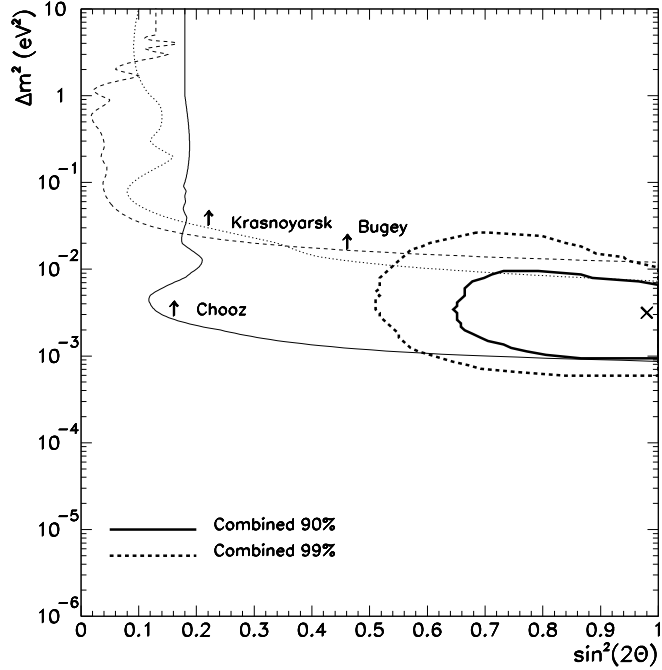,width=0.6\textwidth}}}
\caption{Allowed $\nu_\mu \to \nu_e$ oscillation parameters for all
experiments combined at 90 (solid) and 99 \% CL (dashed) including
Earth matter effects. For comparison we also plot the presently
excluded region from reactor experiments. The cross represents the
best fit point.}
\label{muematall} 
\end{figure}

The allowed $\nu_\mu \to \nu_e$ oscillation parameters for all
experiments combined at 90 and 99 \% CL including Earth matter effects
is shown in \fig{muematall}. Again one can see that by adding the
matter effects the allowed regions lift higher in $\Delta m^2$ than
when matter effects are neglected. We found the best fit point at
($\sin^2 2\theta, \Delta m^2) \sim (0.97, 2.6 \times 10^{-3}$ eV$^2$) where
$\chi^2_{min} = 62.7$ for 40 degrees of freedom. We would like to point
out that at this stage the weight of the experiments with negative 
results (NUSEX and Fr\'ejus) is small enough not to modify the
$\chi^2$ per degree of freedom 
($\chi^2_{min}=54.2$ for 36 degrees of freedom
when these experiments are removed).

It is instructive at this stage to compare the
region determined by the atmospheric neutrino data fit with the
presently excluded region from reactor experiments \cite{bugey}. The
inclusion of the matter effects becomes especially relevant when one
compares with the long baseline reactor neutrino data, such as the
recent data of Chooz \cite{chooz}. One sees that at 90\% CL the \nm to
\ne oscillation channel is ruled out as a solution of the atmospheric
neutrino anomaly.

\subsection{$\nu_\mu \to  \nu_\tau$ channel }

\begin{figure}
\centerline{\protect\hbox{\psfig{file=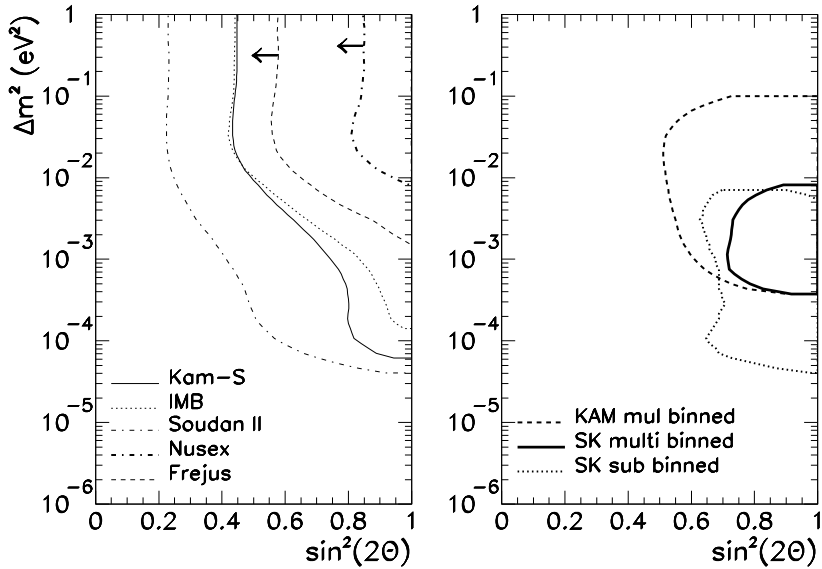,width=0.75\textwidth}}}
\caption{Allowed $\nu_\mu \to  \nu_\tau$ oscillation parameters at 90\% CL
 for each individual experiment.}
\label{mutauie} 
\end{figure}

The results of our $\chi^2$ fit of atmospheric neutrino data obtained
from the data of individual experiments for the $\nu_\mu \to \nu_\tau$
channel are shown in \fig{mutauie}. The allowed regions for each
experiment lie to the right of the corresponding labelled line,
except for the negative Frejus and Nusex experiments.  In the left
figure in \fig{mutauie} we have not included the constraints that
arise from the inclusion of the angular dependence of the data in the
Kamiokande multi-GeV data as well as in the Super-Kamiokande data. 

\begin{figure}
\centerline{\protect\hbox{\psfig{file=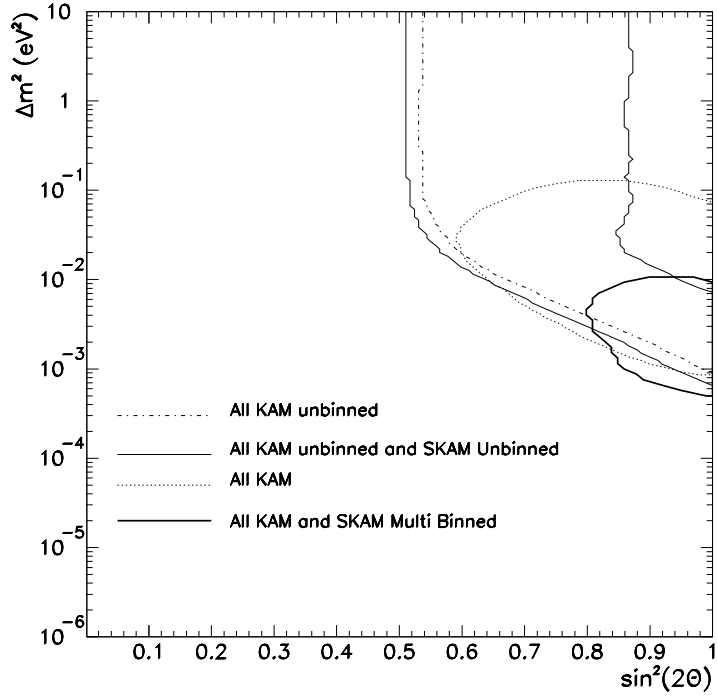,width=0.65\textwidth}}}
\caption{Allowed $\nu_\mu \to  \nu_\tau$ oscillation parameters at 90\% CL
 for Kamiokande and Kamiokande plus Superkamiokande combined.}
\label{mutaukam} 
\end{figure}

It is instructive to compare the results obtained for the Kamiokande
data with those obtained by including the recent Super-Kamiokande
data.  In \fig{mutaukam} we show the allowed $\nu_\mu \to \nu_\tau$
oscillation parameters for Kamiokande and Kamiokande plus
Super-Kamiokande combined. Some features are worth remarking.  For
example, the inclusion of the unbinned Super-Kamiokande data to the
corresponding Kamiokande data leads to the exclusion of large mixing
in the large $\Delta m^2$ region. On the other hand the inclusion of
Super-Kamiokande binned data leads to a substantially smaller region
that obtained from the Kamiokande full data sample, reflecting a
real improvement.

\begin{figure}
\centerline{\protect\hbox{\psfig{file=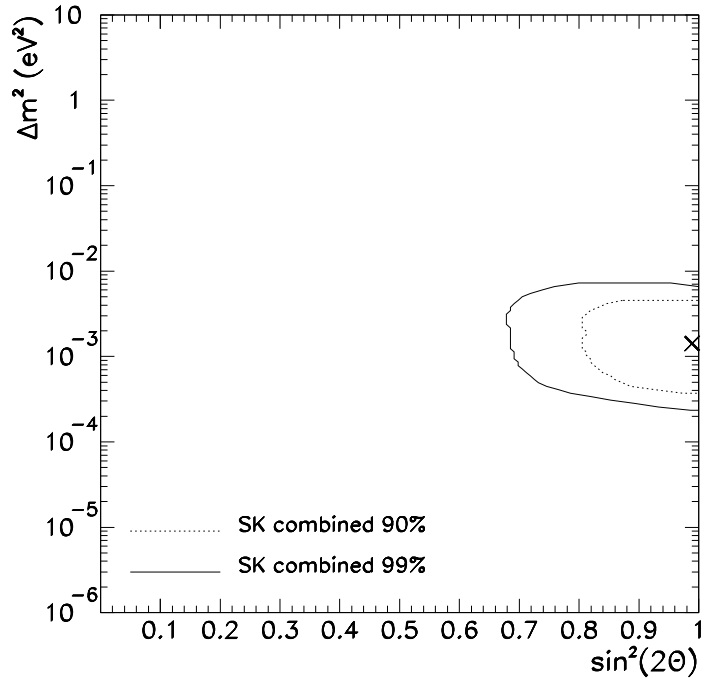,width=0.65\textwidth}}}
\caption{Allowed $\nu_\mu \to  \nu_\tau$ oscillation parameters
 for Superkamiokande combined at 90 and 99 \% CL.}
\label{mutausk} 
\end{figure}

\begin{figure}
\centerline{\protect\hbox{\psfig{file=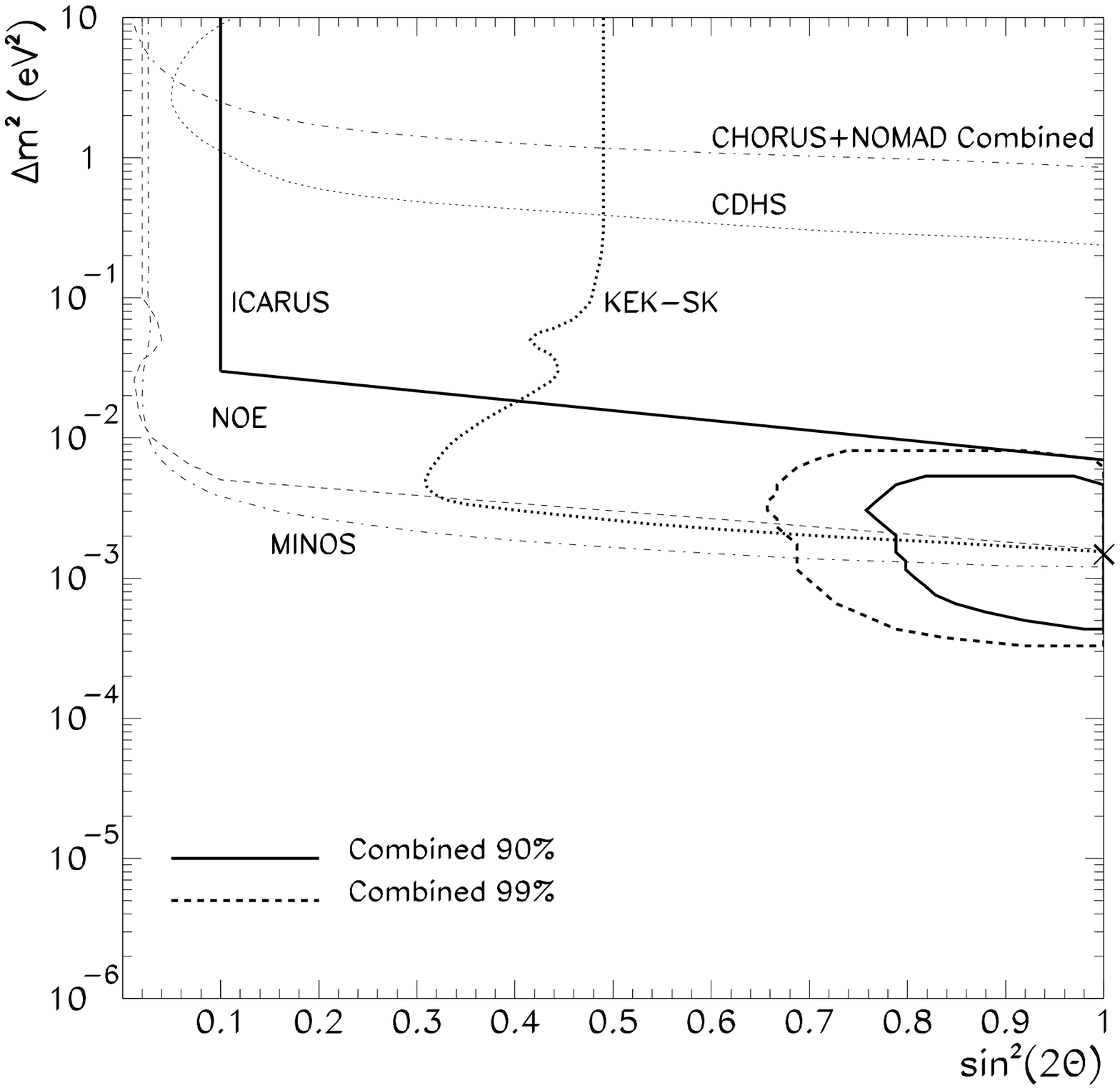,width=0.5\textwidth}}}
\caption{Allowed $\nu_\mu \to  \nu_\tau$ oscillation parameters
 for all experiments  combined at 90 and 99 \% CL. For comparison we also
 display the presently excluded region from accelerator experiments 
 CDHSW and CHORUS+NOMAD and  future long baseline experiments.}
\label{mutauall} 
\end{figure}

In \fig{mutausk} we give the allowed $\nu_\mu \to \nu_\tau$
oscillation parameters for Super-Kamiokande combined at 90 and 99 \%
CL, while in \fig{mutauall} we display the allowed $\nu_\mu \to
\nu_\tau$ oscillation parameters for all experiments combined at 90
and 99 \% CL.  By comparing \fig{mutausk} with \fig{mutauall} one can
see the weight of the Super-Kamiokande data sample in the total data
sample collected by all experiments. 
We find that the best fit points lie at ($\sin^2 2\theta,
\Delta m^2) \sim (1, 1.3 \times 10^{-3} $eV$^2$) with $\chi^2_{min} =
14.4$ for the $20$ degrees of freedom, for Super-Kamiokande only and
($\sin^2 2\theta, \Delta m^2) \sim (1, 1.2 \times 10^{-3} $eV$^2$)
with $\chi^2_{min} = 66.6$ for the 40 degrees of freedom, for all
combined.  The global fit to all experiments is still slightly better
for the $\nu_\mu\rightarrow \nu_e$ channel. However, the difference
between the quality of the fit for both channels is smaller now than
in the pre-Super-Kamiokande era, due to the angular distribution of
Super-Kamiokande multi-GeV which strongly favours the
$\nu_\mu\rightarrow \nu_\tau$ channel.

The result of including the information on the zenith angle
distribution of the events in the \nm to $\nu_\tau$  fit is clearly to cut the
large values of $\Delta m^2$, as can be seen in all figures, namely
\fig{mutauie}, \fig{mutaukam}, \fig{mutausk} and \fig{mutauall}.

One point worth noting is that the inclusion of the new
Super-Kamiokande data produces a downwards shift in the ($\sin^2
2\theta, \Delta m^2)$ region, when compared with pre-Super-Kamiokande
fits.  The importance of the information obtained from the analysis of
the atmospheric neutrino data analysis in relation to the results from
accelerator experiments such as E776 and E531 and CDHSW \cite{CDHSW}
as well as CHORUS plus NOMAD combined limits \cite{dilella} 
and the prospects for the future experiments being discussed at present can be
appreciated in \fig{mutauall}. One sees that the long-baseline
experiments planned at KEK (K2K) \cite{chiaki}, Fermilab (MINOS) 
\cite{minos} and CERN (NOE \cite{noe} and ICARUS \cite{icarus})
fall short in sensitivity to probe the \nm to $\nu_\tau$  oscillation
parameters. This is in contrast with the situation in the
pre-Super-Kamiokande days. From this point of view experiments such as
ICARUS and a re-design of experiments such as MINOS would be desired
in order to enhance their sensitivity in testing the atmospheric
neutrino anomaly.

\section{Discussion and Conclusions}
\label{conclu}

In this paper we have considered the impact of recent experimental
results on atmospheric neutrinos from Super-Kamiokande and Soudan2 as
well as recent theoretical improvements in flux calculations and
neutrino-nucleon cross sections on the determinations of atmospheric
neutrino oscillation parameters, both for the $\nu_\mu \to \nu_\tau$
and $\nu_\mu \to \nu_e$ channels.  The new Super-Kamiokande data cause
a downwards shift in the ($\sin^2 2\theta, \Delta m^2)$ region, when
compared with pre-Super-Kamiokande results.  We have also compared the
results obtained in our fits of atmospheric neutrino data with
previous results, as well as with the constraints following from
laboratory searches for neutrino oscillations, both at accelerators
and reactors. For example we have seen that the $\nu_\mu \to \nu_e$
oscillation hypothesis is barely consistent with the recent negative
result of the Chooz reactor \cite{chooz}. The sensitivity attained in
atmospheric neutrino observations in the $\nu_\mu \to \nu_\tau$
channel is also compared with those of accelerator neutrino
oscillation searches, for example at the present CHORUS and NOMAD as
well as at the future experiments being discussed at
present. Specially interesting from our point of view are the
long-baseline experiments planned at KEK (K2K), Fermilab (MINOS) and
CERN (NOE, ICARUS).  However, due to the lowering of the allowed
($\sin^2 2\theta, \Delta m^2) $ region, it is not clear whether a
re-design is needed in some of these experiments, for example MINOS,
in order to enhance their sensitivity in testing the atmospheric
neutrino anomaly.

Note that, throughout this work we have assumed a rather generous
error in the absolute fluxes of atmospheric neutrinos.  We have
investigated to some extent the effect that a reduced error in the
fluxes would have in the determination of neutrino oscillation
parameters from the present atmospheric neutrino data. We have found
no significant effect in the shape of allowed region when we changed
the assumed error in the fluxes from 30 \% to 20 \%.  However, we have
noticed a somewhat significant effect of a more accurate ratio of
muon-to-electron events. We have found, for example, for
Super-Kamiokande, that when we decrease the error in the
muon-to-electron-type event ratio from 5 \% (10 \%) for unbinned
(binned) data to 3 \% (6\%) the allowed region shrinks by about 10 to 15 \%
 $\sin^2 2\theta$ close to 0.7 or so.  There is hardly any effect in
the $\Delta m^2$ range determination.

\section*{Acknowledgements}

This work was supported by Funda\c{c}\~ao de Amparo \`a Pesquisa do
Estado de S\~ao Paulo (FAPESP), by DGICYT under grant PB95-1077, CICYT
under grant AEN96--1718, and by the TMR network grant ERBFMRXCT960090
of the European Union. H. Nunokawa was supported by a DGICYT
postdoctoral grant, O. Peres by a postdoctoral grant from FAPESP. The
research of T.~S. is supported in part by US Department of Energy
contract DE-FG02-91ER40626.  M.\ C.\ G--G is grateful to the Instituto
de F\'{\i}sica Te\'orica for its kind hospitality. We want to thank
Takaaki Kajita, for providing us with the Kamioka efficiencies and for
useful discussions, Kunio Inoue for providing us with the preliminary
Super-Kamiokande acceptances, Hugh Gallagher, Yoichiro Suzuki, Chiaki
Yanagisawa and Eligio Lisi, for useful correspondence and discussion
and our colleagues Eulogio Oset and S. K. Singh, for useful
discussions on the uncertainties in neutrino nucleus cross sections.

\newpage
\vglue -1cm
\noindent
\section*{Appendix: correlation of errors}

Here we present the errors and correlations  used in our  analysis. In Table
\ref{tab:corre} we display the errors and  correlations $\rho_{\mu e}(A,A)$ for
all the experiments. Data errors and correlations contain the experimental
statistical errors as well as those due to misidentification  as quoted by the
experiments. In order to compute $\sigma_{e}^{theory}$  we take into account
\cite{fogli2}, the flux uncertainty, the MC statistical errors (which depend on
the number of simulated MC events) as well as the cross sections uncertainty.
The flux uncertainty is taken to be $30\%$ whereas MC statistical errors are
estimated under the assumption that the  $\mu$ and $e$-like events follow a
binomial distribution.   Nuclear cross section uncertainties are taken to be 10
\% for  all the experiments except for Soudan2 for when we  used the values,
$7.5 \%$ and $6.4 \%$ for  $e$-like and $\mu$-like events, respectively
\cite{Soudan2}.  

Data errors between different experiments are assumed to be uncorrelated,
\begin{eqnarray}
\rho_{\alpha\alpha}^{data}(A,A) =& 1 \ (\alpha = e, \mu) 
\ \ & \mbox{for\  all}\ \ A, \nonumber \\ 
\rho_{\alpha\beta}^{data}(A,B) = &0 \ (\alpha, \beta = e, \mu)
\ \  &\mbox{if}\ \ A \neq  \  B, \nonumber
\end{eqnarray}
while the theory correlations
between different experiments (i.e., for $A \neq B$) are obtained as follows, 
\begin{displaymath}
\rho_{\alpha\beta}^{theory} (A,B) 
= \rho_{\alpha\beta}^{flux}\times \frac{ \sigma_\alpha^{flux} 
\sigma_\beta^{flux} }
{\sigma_\alpha^{theory}(A)\ \sigma_\alpha^{theory}(B)}
\ \ \ \mbox{if}\ \alpha \neq  \beta, 
\end{displaymath}
\begin{displaymath}
\rho_{\alpha\beta}^{theory} (A,B) 
= \rho_{\alpha\beta}^{flux} \times \frac{ \sigma_\alpha^{flux} 
\sigma_\beta^{flux} }
{\sigma_\alpha^{theory}(A)\ \sigma_\beta^{theory}(B)}
\ \ \ \mbox{if}\ \alpha =  \beta, 
\end{displaymath}
where 
$\sigma_e^{flux} =\sigma_\mu^{flux} = 30$ \% and 
$\rho_{\alpha\beta}^{flux} = 1.000$  for $\alpha =  \beta$. For
$\alpha \neq \beta$, we use $\rho_{\alpha\beta}^{flux}= 0.986$ (0.944) 
as determined from the relation 
\begin{displaymath}
{( \sigma_{\mu/e}^{flux} )}^2 = {(\sigma_\mu^{flux})}^2 + 
{(\sigma_e^{flux})}^2 - 2\rho_{\mu e} (\sigma_\mu^{flux})
(\sigma_e^{flux})
\end{displaymath}
after imposing that the uncertainty in the flavor ratio, 
$\sigma_{\mu/e}^{flux}=5 \% (10 \%)$ for unbinned (binned) case 
\cite{fogli2}. Furthermore we assume that there is no correlation 
between the sub-GeV and multi-GeV data.

We note that for both sub-GeV and multi-GeV data, 
in general, $\rho_{\alpha\beta}^{theory} (A,B) $ 
is not symmetric under the exchange of the flavor labels 
$\alpha$ and $\beta$ or the experimental labels , i.e.,   
\begin{eqnarray}
\rho_{\alpha\beta}^{theory}(A,B) \neq \rho_{\beta\alpha}^{theory}(A,B) &
\ \ \ &\mbox{if}\ \alpha \neq \beta, \nonumber \\
\rho_{\alpha\beta}^{theory}(A,B) \neq \rho_{\alpha\beta}^{theory}(B,A) &
\ \ \ &\mbox{if}\ A \neq B,\nonumber
\end{eqnarray}
but it is symmetric under simultaneous exchange of both 
kinds of labels $\alpha$, $\beta$ and $A$, $B$:
\begin{displaymath}
\rho_{\alpha\beta}^{theory}(A,B) = \rho_{\beta\alpha}^{theory}(B,A).\nonumber
\end{displaymath}

%
\newpage 
\begin{table}[h]
\begin{center}
\begin{tabular}{||l|l||}
Experiment &  $R_{\mu/e}/R^{MC}_{\mu/e}$ \\\hline
Super-Kamiokande sub-GeV & $0.635 \pm 0.035 \pm 0.053 $\\
Super-Kamiokande multi-GeV &  $0.604 \pm 0.065 \pm 0.065$\\
Soudan2  & $0.61 \pm 0.14 \pm 0.07$ \\
IMB & $0.55 \pm 0.11$ \\
Kamiokande sub-GeV & $0.6 \pm 0.09$ \\
Kamiokande multi-GeV & $0.59 \pm 0.1 $\\
Fr\'ejus & $1.06 \pm 0.23$ \\
Nusex  & $0.96 \pm 0.3$\\
\end{tabular}
\end{center}
\caption[Tab]{Results from the atmospheric neutrino experiments}
\label{tab:data}
\end{table}
\begin{table}[h]
\begin{center}
\begin{tabular}{||l|l|l||l|l|l||}
 & $\frac{N_{\mu}^{MC}}{N_{e}^{MC}}$
& $\frac{N_{\mu}^0}{N_{e}^0}$ & & 
$\frac{N_{\mu}^{MC}}{N_{e}^{MC}}$
& $\frac{N_{\mu}^0}{N_{e}^0}$ 
\\ \hline\hline
 Fr\'ejus        & 1.9 & 1.8 & Super-Kamiokande (sub-GeV)& 1.6 & 1.6  \\ \hline
 Kamiokande (sub-GeV)  & 1.55 & 1.6 & Bin1 & 1.7 & 1.6 \\ \hline
 IMB            & 1.1 & 1.1  & Bin2 & 1.6 & 1.5 \\ \hline
 Soudan2        & 1.05 & 1.1 & Bin3 & 1.5 & 1.5  \\ \hline
 Nusex         & 1.9  & 1.8 & Bin4 & 1.5 & 1.6   \\ \hline
 Kamiokande (multi-GeV)& 2.3 & 2.4  & Bin5 & 1.7 & 1.5\\ \hline
 Bin1 & 3.1 & 3.1  &  Super-Kamiokande (multi-GeV)& 3.2  &  3.0 \\ \hline
 Bin2 & 2.4 & 2.4  &   Bin1 & 3.8 & 3.4\\ \hline
 Bin3 & 2.1 & 2.0   &  Bin2 & 2.8 & 2.8\\ \hline
 Bin4 & 2.4 & 2.4  &   Bin3 & 3.2 & 2.8\\ \hline
 Bin5 & 3.2 & 3.2  &   Bin4 & 2.9 & 2.8 \\ \hline
      &     &      &   Bin5 & 4.2 & 3.5   \\
\end{tabular}
\end{center}
\caption[Tab]{Our predictions for the ratio ($N_{\mu}^0/N_{e}^0$)  in the
absence of oscillations compared to the MC expectations 
($N_{\mu}^{MC}/N_{e}^{MC}$) from each experimental group.} 
\label{tab:our}
\end{table}
\newpage
\begin{table}[h]
\begin{center}
\begin{tabular}{||c||cc||}
Experiment   &\  \  \  $\chi^2$ & C. L. (\%)  \\
\hline
 Fr\'ejus &0.56 & 24.4\\
IMB & 8.4 & 98.5 \\
Soudan2 & 5.7 &94.2\\
Nusex & 0.39 &17.7\\
 Kamiokande sub-GeV & 12.5& 99.8\\
 Kamiokande multi-GeV unbinned& 8.7 &98.7\\
 Kamiokande multi-GeV binned& 18.2 & 94.8\\
Super-Kamiokande sub-GeV unbinned& 21.5 &99.7\\
Super-Kamiokande sub-GeV binned& 27.2 &100.0\\
Super-Kamiokande multi-GeV unbinned & 10. &99.3\\
Super-Kamiokande multi-GeV binned & 27.9 &99.8\\
\end{tabular}
\end{center}
\caption[Tab]{Values of $\chi^2$ and confidence level for each
experiment in the absence of oscillations. For unbinned data the
number of degrees of freedom is 2 while for combined binned data is
10.}
\label{tab:chi2}
\end{table} 
 
\newpage
\vskip -2.0cm
{\def\arraystretch{1.3}
\begin{table}[h]
\begin{center}
\begin{tabular}{||c||ccc||ccc||}
Experiment ($A$) & $\sigma_{\mu}^{data}$ & $\sigma_{e}^{data}$ 
 (\%)  &$\rho_{\mu e}^{data}(A,A)$ 
& $\sigma_{\mu}^{theory}$  & $\sigma_{e}^{theory}$ (\%)  
  & $\rho_{\mu e}^{theory}(A,A)$\\ 
\hline\hline
Fr\'ejus & 10.5 & 17.9 & $-$0.021 & 31.7 &31.9 & 0.951 \\
Kam sub-GeV & 7.1 & 7.0 &$-$0.081 & 31.7 & 31.8   & 0.975 \\
IMB & 8.9 &7.5 &$-$0.374 & 36.1 & 36.1   & 0.947 \\
Nusex & 18.4 & 27.2 &$-$0.050 & 31.7 & 31.9   & 0.950 \\
Soudan2 & 13.5 & 11.0 &$-$0.168 & 30.8 & 31.1  & 0.960 \\
Super-Kam sub-GeV (unbinned)
& 4.9 &4.9 & $-$0.042 & 31.6 & 31.7  & 0.978 \\ \hline
Super-Kam sub-GeV bin1 
& 9.4 & 8.3 & $-$0.013 & 31.7 & 31.8  & 0.936 \\ 
Super-Kam sub-GeV bin2 
& 9.0 & 9.4 & $-$0.012 & 31.7 & 31.8  & 0.935 \\ 
Super-Kam sub-GeV bin3 
& 9.0 & 8.4 & $-$0.013 & 31.7 & 31.8  & 0.936 \\ 
Super-Kam sub-GeV bin4 
& 8.6 & 9.1 & $-$0.013 & 31.7 & 31.8  & 0.936 \\ 
Super-Kam sub-GeV bin5
& 8.3 & 9.8 & $-$0.012 & 31.7 & 31.8  & 0.935 \\ \hline \hline 
 Kam multi-GeV  (unbinned) 
& 9.6 & 11.0 & $-$0.038 & 31.7 & 32.0  &  0.965\\ \hline
 Kam multi-GeV bin1 
& 24.0 & 22.2 &$-$0.008 & 31.8 & 33.9   & 0.840 \\ 
Kam multi-GeV bin2 
& 22.8 & 23.3 &$-$0.008 & 31.9 & 32.9   & 0.869 \\ 
 Kam multi-GeV bin3 
& 18.2 & 19.1 &$-$0.012  & 31.8 & 32.6  & 0.889 \\ 
 Kam multi-GeV bin4 
& 18.8 & 22.8 &$-$0.009 & 31.9 & 32.8   & 0.878 \\ 
 Kam multi-GeV bin5 
& 17.2 & 33.6 &$-$0.007 & 31.9 & 33.8   & 0.838 \\  \hline
Super-Kam multi-GeV (unbinned)
& 7.6 & 9.5 & $-$0.056 & 31.6 & 31.9  & 0.972 \\  \hline
Super-Kam multi-GeV bin1 
& 17.9 & 21.9  &$-$0.010 & 31.7 & 32.4  & 0.911\\  
Super-Kam multi-GeV bin2 
& 17.4 & 18.1 & $-$0.013 & 31.7 &32.1  & 0.920 \\  
Super-Kam multi-GeV bin3 
& 12.4 & 18.9   &$-$0.017 & 31.7 & 32.1  & 0.923 \\  
Super-Kam multi-GeV bin4 
& 12.2 & 17.1  &$-$0.019 & 31.7 & 32.1 & 0.919 \\  
Super-Kam multi-GeV bin5 
& 12.7 & 19.8 &$-$0.016 & 31.7 & 32.4  & 0.909 \\
\end{tabular}
\end{center}
\caption[Tab]{Errors and correlation for both 
observed data and theory (MC) samples.}
\label{tab:corre}
\end{table}
}

%

\vs 2cm

\begin{references}

\bibitem{review} For reviews, see for e.g., 
T.K. Gaisser, in Neutrino '96, Proc. of the 17th International Conference
on Neutrino Physics and Astrophysics, Helsinki, Finland, 
edited by K. Enquist, K. Huitu and J. Maalampi (World Scientific, 1997),
p. 211; T.K. Gaisser, F. Halzen and T. Stanev, Phys. Rep. 258, 174 (1995);
T. Stanev, Proceedings in TAUP '95, Toledo, Spain, \nps{48}{96}{165}; 
T. Kajita, in {\it Physics and Astrophysics of Neutrinos}, ed. by 
M. Fukugita and A. Suzuki (Springer-Verlag,Tokyo, 1994), p. 559; 
E. Kh. Akhmedov, in {\it Cosmological Dark Matter}, Proceedings of 
the International School on Cosmological Dark Matter, Valencia, 
ed.  A. Perez and J.W.F.Valle 
(World Scientific, Singapore, 1994, ISBN 981-02-1879-6), p. 131. 

\bibitem{stanev}
T. Stanev in ref. \cite{review}.

\bibitem{Barish}
B. Barish, in proceedings of Int. Workshop on Elementary
Particle Physics: Present and Future, ed. A. Ferrer and J. W. F. Valle 
(World Scientific, 1996, ISBN-981-02-2554-7), p. 400.

\bibitem{kamisub} Kamiokande Collaboration, H. S. Hirata {\sl et al.},
Phys. Lett. {\bf B205}, 416 (1988) and Phys. Lett. {\bf B280}, 146  (1992);
Kamiokande Collaboration, K. Kaneyuki {\sl et al.}, in 
Proceedings of the XIII Moriond Workshop, 
{\sl Perspectives in Neutrinos, Atomic Physics and 
Gravitation}, p. 211, ed. by J. Tran Thanh Van {\sl et al.}, 
Villars sur Ollon, Switzerland, Editions Frontieres, 1993. 


\bibitem{kamimul} 
Kamiokande Collaboration, Y. Fukuda {\sl et al.}, Phys.  Lett. {\bf
B335}, 237 (1994).

\bibitem{IMB} 
IMB Collaboration, D. Casper {\sl et al.}, Phys. Rev. Lett.  {\bf 66},
2561 (1991); R. Becker-Szendy {\sl et al.}, Phys. Rev. {\bf D46}, 3720
(1992); IMB Collaboration, D. Keilczewska {\sl et al.}, in 
Proceedings of the XIII Moriond Workshop, 
{\sl Perspectives in Neutrinos, Atomic Physics and 
Gravitation}, p. 219, ed. by J. Tran Thanh Van {\sl et al.}, 
Villars sur Ollon, Switzerland, Editions Frontieres, 1993. 

\bibitem{frejus} Fr\'ejus Collaboration, Ch.\ Berger {\sl et al.},
Phys.\ Lett.\ {\bf B227},  489 (1989).

\bibitem{nusex}  NUSEX Collaboration,
M.\ Aglietta {\sl et al.}, Europhys.\ Lett.\ {\bf 8}, 611  (1989).

\bibitem{Soudan2} Soudan Collaboration, W.\ W.\ M\ Allison {\sl et
al.}, Phys.\ Lett.\ {\bf B391}, 491 (1997); Soudan Collaboration, Hugh
Gallagher talk at ``XVI International Workshop on Weak Interactions
and Neutrinos'', Capri, Italy, (June, 1997). Soudan Collaboration,
T. Kafka talk at TAU97, September 7-11, 1997 - Laboratori Nazionali
del Gran Sasso, Assergi (Italy), hep/ph 9712281.

\bibitem{agrawal} V.\ Agrawal {\sl et al.}, Phys.\ Rev.\ {\bf D53},
1314 (1996) and {\sl An Improved Calculation of the Atmospheric
Neutrino Flux}, T.K.Gaisser and T.Stanev, in {Proc. 24th ICRC},
Vol. 1, p. 694 (Rome 1995).

\bibitem{volkova} \ L.\ V.\ Volkova, Sov.\ J.\ Nucl.\ Phys.\ {\bf 31},
784 (1980) .

\bibitem{chooz} 
M. Apollonio {\sl et al.}, CHOOZ Collaboration,
preprint, hep-ex/9711002.

\bibitem{BGS} G.\,Barr, T.\,K.\,Gaisser and T.\,Stanev,
Phys. Rev. {\bf D\,39} (1989) 3532 and Phys.\ Rev.\ {\bf D38}, 85
(1988)

\bibitem{LSgeomag} P.\ Lipari and T.\ Stanev, in {Proc. 24th ICRC},
 Vol. 1, (Rome 1995)

\bibitem{HKHM} 
M.\,Honda, T. \ Kajita, K.\,Kasahara, and S.\,Midorikawa, 
Phys. Rev. {\bf D\,52} 4985 (1995);
M.\,Honda, K.\,Kasahara, K.\,Hidaka and S.\,Midorikawa, Phys.
Lett. B {\bf 248} 193 (1990).

\bibitem{Getall} T.\ K.\ Gaisser {\it et al., Phys. Rev. D}{\bf 54},
 5578 (1996)

\bibitem{BN} 
E.\,V.\,Bugaev and V.\,A.\,Naumov, Phys. Lett. B {\bf 232} (1989)
391.

\bibitem{Circella} M.\ Circella {\it et al.,} in 
{\it Proc. 25th ICRC}, Vol. 7, p. 117 (Durban 1997)
 
\bibitem{pathlength} 
T. K. Gaisser and T. Stanev, preprint BRI-97-28, astro-ph/9708146. 

\bibitem{paolo} 
P. Lipari, M. Lusignoli, and F. Sartogo, \prl{74}{95}{4384}

\bibitem{smith} C.\ L.\ Smith, Phys.\
Rep.\ {\bf C3}, 261  (1972).

\bibitem{cross} G.\ L.\ Fogli and G.\ Nardulli, Nucl.\ Phys.\ {\bf B160}, 116
(1979); M.\ Nakahata {\sl et al.}, J.\ Phys.\ Soc.\ Japan {\bf 55}, 3786
(1986).

\bibitem{kajita} T.\ Kajita, private communication, see also
T. Kajita in ref. \cite{review}. 

\bibitem{inoue} K.\ Inoue, private communication. 

\bibitem{fogli2}
G.\ L.\ Fogli, E.\ Lisi,  Phys.\ Rev.\  {\bf D52}, 2775 (1995)

\bibitem{msw} S.\ P.\ Mikheyev and A.\ Yu.\ Smirnov, Yad.\
Fiz.\ {\bf 42}, 1441 (1985); L.\ Wolfenstein, Phys.\ Rev.\ {\bf D17}, 2369
(1985).

\bibitem{lisi}
E.\ Lisi and D.\ Montanino,  Phys.\ Rev.\  {\bf D56}, 1792 (1997).

\bibitem{krasnoyarsk} 
G. S. Vidyakin {\sl et al.}, JETP Lett. {\bf 59}, 364 (1994). 

\bibitem{bugey} 
B. Achkar {\sl et al.}, Nucl. Phys. {\bf B424}, 503 (1995).

\bibitem{CDHSW} CDHSW Collaboration, F. Dydak {\sl et al.}, Phys.
Lett.  {\bf B134}, 281 (1984);\\ E776 Collaboration, L.\ Borodvsky
{\sl et al.}, Phys.\ Rev.\ Lett.\ {\bf 68}, 274 (1992);\\ E531
Collaboration, Phys.\ Rev.\ Lett.\ {\bf 57}, 2898 (1986).

\bibitem{chorus} CHORUS Collaboration, N. Armenise {\sl et al.},
CERN-SPSC/90-42 (1990).

\bibitem{nomad} NOMAD Collaboration, P. Astier {\sl et al.},
CERN-SPSLC/91-21 (1991), CERN-SPSLC/91-48 (1991), SPSLC/P261 Add. 1 (1991).

\bibitem{dilella} 
L. Di Lella, talk at TAUP97, September 7-11, 1997 - Laboratori 
Nazionali del Gran Sasso, Assergi.

\bibitem{chiaki} KEK-SK Collaboration, C. Yanagisawa, talk at
International Workshop on Physics Beyond The Standard Model: from
Theory to Experiment, Valencia, Spain, October 13-17, 1997, to appear
in the proceedings, ed. by I. Antoniadis, L. Ibanez and J. W. F. Valle
(World Scientific, 1998)

\bibitem{minos} MINOS Collaboration, NuMI-L-63 MINOS proposal, Dave Ayres 
{\sl et al.}, and  the homepage http://www.hep.anl.gov/NDK/HyperText/numi.html.

\bibitem{noe} NOE Collaboration, M . Ambrosio {\sl et al.}, 
Nucl Instr. Meth. A363, 604 (1995) and the
homepage http://www.na.infn.it/SubNucl/accel/noe/noe.html

\bibitem{icarus} ICARUS Collaboration http://www.aquila.infn.it/icarus/ 
and ''A first 600 ton ICARUS detector installed at the Gran Sasso Laboratory",
Addendum to Proposal by the ICARUS Collaboration, LNGS - 95/10 (1995).

\end{references}
\end{document}